\documentclass{emulateapj}
\usepackage{threeparttable}
\usepackage{bm}
\usepackage{chngpage}
\usepackage{graphicx}
\usepackage{mathrsfs}

\shorttitle{Premerger EM Emission of Black Hole-Neutron Star Binaries}
\shortauthors{Dai}

\begin{document}

\title{Inspiral of a Spinning Black Hole--Magnetized Neutron Star Binary: Increasing Charge and Electromagnetic Emission}

\author{Z. G. Dai\altaffilmark{1,2}}
\affil{\altaffilmark{1}School of Astronomy and Space Science, Nanjing University, Nanjing 210023, China; dzg@nju.edu.cn}
\affil{\altaffilmark{2}Key Laboratory of Modern Astronomy and Astrophysics (Nanjing University), Ministry of Education, Nanjing, China}

\begin{abstract}
The mergers of black hole (BH)--neutron star (NS) binaries have been one of the most interesting topics in astrophysics, because such events have been thought to possibly produce multimessenger signals including gravitational waves and broadband electromagnetic (EM) waves. In this paper, we investigate EM emission from the inspiral of a binary composed of a spinning BH and a magnetized NS. Observationally, the BH is usually more massive than $\sim7M_\odot$ and the NS has a mass $\simeq 1.4M_\odot$. During the inspiral of such a binary, the BH will accumulate more and more charges based on the charging scenario of Wald, even though the BH will eventually swallow the NS whole inevitably. We calculate the emission luminosities and energies through three energy dissipation mechanisms: magnetic dipole radiation, electric dipole radiation, and magnetic reconnection. We show that magnetic dipole radiation due to the spin of the increasingly charged BH and magnetic reconnection in between the BH and the NS could be most significant at the final inspiral stage. We find that if the BH is rapidly rotating and the NS is strongly magnetized, these mechanisms would lead to a detectable EM signal (e.g., a short-duration X-ray transient).
\end{abstract}


\section{Introduction}
Since the mergers of black hole--black hole (BH--BH) binaries \citep{Abbott16,Abbott18a} and neutron star--neutron star (NS--NS) binaries \citep[i.e., GW170817,][]{Abbott17a,Abbott18b} were discovered by the advanced LIGO/Virgo gravitational wave (GW) detectors, observations of BH--NS mergers have become one of the most interesting topics in astrophysics. This is because it has been widely argued that BH--NS mergers not only lead to GW events \citep{Val00,Shi08,Shi09,Kyu10,Kyu11,Cho13,Tag14,Pan14,Kum15,Pan15} but also perhaps generate broadband electromagnetic (EM) signals, e.g., short-duration gamma-ray bursts and multiwavelength afterglows driven by ultrarelativistic jets \citep[][for a review]{Pac91,Narayan92,Moch93,Janka99,Davies05,Berger14}, kilonovae powered by radioactive element decay in highly anisotropic ejecta \citep{Li98,Kyu13,Kyu15,Kaw16,Huang18}, and radio transients from ejecta-medium interactions \citep{Nakar11}, similar to NS--NS mergers \citep[for a review, see][]{Bai17}. In addition, BH--NS mergers would also be used to measure the Hubble constant \citep{Vit18}, as in the case of GW170817 \citep{Abbott17b}.

For a BH--NS merger, possible EM signals mentioned above are highly dependent of the properties of tidally disrupted ejecta and fallback matter. In principle, if the premerger BH--NS mass ratio ($q$) is so low that the disruption radius that equals to a factor of $q^{1/3}$ times the NS radius is larger than the BH's Schwarzschild radius, then a fraction of NS matter would be possibly disrupted tidally during the merger and EM emission would be produced. On the other hand, if $q$ is high enough, then the NS whole would plunge into the BH. In this case, matter is neither ejected nor does it fall back during the merger, so no EM emission is expected. A simple analysis suggests that the critical value of $q$ is $\sim3.6(M_{\rm *}/1.4M_\odot)^{-3/2}(R_{\rm *}/10^6\,{\rm cm})^{3/2}$ (where $M_{\rm *}$ and $R_{\rm *}$ are the NS's mass and radius,\footnote{All of the quantities marked with a subscript ``$\ast$'' (or ``$\bullet$'') indicate those of an NS (or BH) in this paper.} respectively); a detailed study shows that this critical value should be $\sim 5$ \citep{Shi09}. This implies that if the BH mass $\lesssim7M_\odot$ for $M_{\rm *}\simeq1.4M_\odot$, a fraction of NS matter would be tidally disrupted at the inspiral stage; if the BH mass $\gtrsim7M_\odot$, however, the NS as a whole would possibly plunge into the BH.

The observations of BH--BH mergers by aLIGO/Virgo indicate that all of the BH masses are $\gtrsim7M_\odot$ in these systems \citep{Abbott18a}, while mass measurements on galactic BHs show that most of the BHs in X-ray transients and high-mass X-ray binaries have masses above this critical mass \citep{Cas14,McC14}. Therefore, it would be expected that all of the BH--NS binaries may have a mass ratio $q\gtrsim5$ for $M_{\rm *}\simeq1.4M_\odot$ and their mergers could not eject NS matter, so such events would be unable to produce any EM emission. Two scenarios have been proposed to generate EM emission. First, if the NS in a BH--NS binary is strongly magnetized, the BH will cross magnetic field lines around the NS at the inspiral stage, possibly leading to instantaneous acceleration of electrons to relativistic energies and coherent curvature radiation, further resulting in a fast radio burst and even a fireball \citep{Ming15,DOr16}. Second, if the BH is charged initially and constantly, then both electric dipole radiation and magnetic dipole radiation before binary merger would produce interesting EM signals, as discussed by \cite{Zhang19} following earlier studies of EM emission from charged BH--BH mergers \citep{Zhang16,Liu16}. In this scenario, however, the EM force is extremely strong so that the BH will discharge instantaneously through absorbing opposite charges swiftly from an ionized interstellar medium \citep{Levin18}.

In this Letter, we investigate EM emission from the inspiral of a binary composed of a rapidly rotating BH and a strongly magnetized NS, in which the BH, immersed in the NS's magnetic field, will accumulate more and more charges at the inspiral stage of the binary based on the charging scenario of \cite{Wald74}. We discuss some implications of this scenario in detail. This paper is organized as follows. We first discuss Wald's scenario and two magnetic dipole moments which originate from the increasingly charged BH's spin and  binary rotation, respectively, in Section \ref{cbh}. We next discuss three energy dissipation mechanisms to generate EM emission in Section \ref{edm}, and provide order-of-magnitude estimates of their emission luminosities and energies as well as the detectable merger event rate in the X-ray band in Section \ref{est}. Finally, we summarize our conclusions in Section \ref{con}.

\section{Increasingly Charged BH}\label{cbh}
We consider a rotating BH with mass $M_{\rm \bullet}$ of order $\sim10M_\odot$ and angular momentum $J_{\rm \bullet}=a_\bullet(GM_{\rm \bullet}^2/c)$, where $a_\bullet\gtrsim0.5$ is the BH's dimensionless spin parameter. We also consider a magnetized NS with mass $M_{\rm \ast}\simeq1.4M_\odot$ and surface magnetic dipole field $B_{\rm s,*}\gtrsim10^{12}\,$G. The two compact objects form a binary. Such a binary system could arise from an X-ray binary with a high-spin BH \citep{McC11,McC14}. The other possibility is that a wandering NS is captured by an isolated high-spin BH originating from an earlier BH--BH merger or some other astrophysical processes, leading to a binary. If the two objects in the BH--NS binary are assumed to move in a nearly circular orbit of radius $r$ about each other, then their respective distances from the center of mass are given by $r_{\rm \bullet}=r(\mu/M_{\rm \bullet})=r(M_{\rm *}/M)$ and $r_{\rm *}=r(\mu/M_{\rm *})=r(M_{\rm \bullet}/M)$, where $M\equiv M_{\rm \bullet}+M_{\rm *}$ is the total mass and $\mu\equiv M_{\rm \bullet}M_{\rm *}/M$ is the reduced mass.

The GW luminosity, total energy, and angular velocity of the binary at its inspiral stage are approximated by \citep{Shapiro83}
\begin{equation}
L_{\rm GW}\equiv -\frac{dE}{dt}=\frac{2^5}{5}\frac{G^4}{c^5}\frac{M^3\mu^2}{r^5}\label{gw},
\end{equation}
\begin{equation}
E=-\frac{1}{2}\frac{GM\mu}{r}\label{totE},
\end{equation}
and from Kepler's III law,
\begin{equation}
\Omega=\left(\frac{GM}{r^3}\right)^{1/2}=\frac{1}{\sqrt{2\kappa}}\frac{c}{r}\left(\frac{R_{\rm S,\bullet}}{r}\right)^{1/2}\label{Omega},
\end{equation}
where $c$ is the speed of light, $\kappa\equiv M_\bullet/M$, and $R_{\rm S,\bullet}=2GM_{\rm \bullet}/c^2=2.96\times10^6(M_{\rm \bullet}/10M_\odot)\,$cm is the BH's Schwarzschild radius. According to Equations (\ref{gw}) and (\ref{totE}), we thus obtain the first and second derivatives of $r$,
\begin{equation}
\dot{r}=-\frac{2^6}{5}\frac{G^3}{c^5}\frac{M^2\mu}{r^3}=-\frac{2^3}{5\kappa}\left(\frac{R_{\rm S,\bullet}}{r}\right)^2\left(\frac{R_{\rm S,*}}{r}\right)c\label{1dr},
\end{equation}
and
\begin{eqnarray}
\ddot{r}=-\frac{3\times2^6}{5^2\kappa^2}\frac{c^2}{r}\left(\frac{R_{\rm S,\bullet}}{r}\right)^4\left(\frac{R_{\rm S,*}}{r}\right)^2\label{2dr},
\end{eqnarray}
where $\mu=\kappa M_*$ is used and $R_{\rm S,*}=2GM_{\rm *}/c^2=4.15\times10^5(M_{\rm *}/1.4M_\odot)\,$cm is the NS's Schwarzschild radius.

For simplicity, we further assume that both the BH's spin and the NS's magnetic dipole axis are parallel (or antiparallel) with the angular momentum of the binary. Therefore, the magnetic field strength at the BH's position is given by $B_{\rm *}=B_{\rm s,*}(R_{\rm *}/r)^3$. \cite{Wald74} first pointed out that this magnetic field will induce a radial electric field and accretion of charged particles onto the BH. \cite{Levin18} argued that the charging timescale is much shorter than the binary inspiral timescale before merger, suggesting that the BH charging will take place instantaneously. \cite{Wald74} got an equilibrium value of the charge quantity through
\begin{eqnarray}
Q_{\rm W} & = & \frac{2G}{c^3}J_{\rm \bullet}B_{\rm *}=\frac{1}{2}a_\bullet R_{\rm S,\bullet}^2B_{\rm *}\nonumber\\
& = & (4.4\times 10^{24}\,{\rm e.s.u.})a_\bullet\left(\frac{M_{\rm \bullet}}{10M_\odot}\right)^2\left(\frac{B_{\rm *}}{10^{12}{\rm G}}\right)\label{QW}.
\end{eqnarray}

Rotation of this charged BH along its spin axis leads to the first magnetic dipole moment, whose expression is given by \citep{Wald74,Liu16,Levin18}
\begin{eqnarray}
|{\bf m}_{\rm \bullet,1}| & = & \frac{J_{\rm \bullet}}{M_{\rm \bullet}c}Q_{\rm W}=\frac{1}{4}a_\bullet^2 R_{\rm S,\bullet}^3B_{\rm *}\nonumber \\
& = & 6.5\times 10^{30}a_\bullet^2\left(\frac{M_{\rm \bullet}}{10M_\odot}\right)^3\left(\frac{B_{\rm *}}{10^{12}{\rm G}}\right){\rm G\,cm}^3\label{mBH1},
\end{eqnarray}
which would give rise to a magnetic dipole field of the BH whose strength at any radius $r'$ is
\begin{eqnarray}
B_{\rm \bullet} & = & 6.5\times 10^9a_\bullet^2\left(\frac{M_{\rm \bullet}}{10M_\odot}\right)^3\nonumber\\
& & \times\left(\frac{B_{\rm *}}{10^{12}{\rm G}}\right)\left(\frac{r'}{10^7{\rm cm}}\right)^{-3}{\rm G}\label{BBH}.
\end{eqnarray}
On the other hand, rotation of the charged BH along the center of mass of the binary leads to the second magnetic dipole moment of order \citep{Zhang16,Zhang19}
\begin{eqnarray}
|{\bf m}_{\rm \bullet,2}| & = & \frac{\pi r_{\rm \bullet}^2}{c}\frac{Q_{\rm W}}{P}\nonumber\\
& = & \frac{1}{4\sqrt{2\kappa}}\left(\frac{M_{\rm *}}{M}\right)^2\left(\frac{r}{R_{\rm S,\bullet}}\right)^{1/2}a_\bullet R_{\rm S,\bullet}^3B_*\label{mBH2},
\end{eqnarray}
where $P=2\pi/\Omega$ is the orbital period of the binary. From Equations (\ref{mBH1}) and (\ref{mBH2}), the ratio of the two magnetic dipole moments is
\begin{eqnarray}
\frac{|{\bf m}_{\rm \bullet,2}|}{|{\bf m}_{\rm \bullet,1}|}=\frac{1}{\sqrt{2\kappa}a_\bullet}\left(\frac{M_{\rm *}}{M}\right)^2\left(\frac{r}{R_{\rm S,\bullet}}\right)^{1/2}\label{ratio}.
\end{eqnarray}
For $a_\bullet\gtrsim 0.5$, $M_{\rm *}\ll M_{\rm \bullet}$ and $r\sim R_{\rm S,\bullet}$, this ratio is much less than unity, implying that $|{\bf m}_{\rm \bullet,2}|$ is negligible as compared with $|{\bf m}_{\rm \bullet,1}|$.

\section{Energy Dissipation Mechanisms}\label{edm}
We next discuss three energy dissipation mechanisms that are able to produce premerger EM emission.

\subsection{Magnetic Dipole Radiation}

Owing to time-changing $|{\bf m}_{\rm \bullet,1}|$ in Equation (\ref{mBH1}), we derive the magnetic dipole radiation luminosity,
\begin{eqnarray}
L_{\rm MDR,1}=\frac{2|\ddot{{\bf m}}_{\rm \bullet,1}|^2}{3c^3}=\frac{6|{\bf m}_{\rm \bullet,1}|^2}{c^3}\left(\frac{4\dot{r}^2-r\ddot{r}}{r^2}\right)^2\label{lm1}.
\end{eqnarray}
On the other hand, following \cite{Zhang16,Zhang19}, we obtain the magnetic dipole radiation luminosity due to varying $|{\bf m}_{\rm \bullet,2}|$ in Equation (\ref{mBH2}),
\begin{eqnarray}
L_{\rm MDR,2}=\frac{2|\ddot{{\bf m}}_{\rm \bullet,2}|^2}{3c^3}=\frac{25|{\bf m}_{\rm \bullet,2}|^2}{6c^3}
\left(\frac{7\dot{r}^2-2r\ddot{r}}{2r^2}\right)^2\label{lm2}.
\end{eqnarray}
According to Equation (\ref{ratio}), therefore, we find $L_{\rm MDR,1}\gg L_{\rm MDR,2}$ for $a_\bullet\gtrsim 0.5$, $M_{\rm *}\ll M_{\rm \bullet}$ and $r\sim R_{\rm S,\bullet}$. Insertion of Equations (\ref{1dr}) and (\ref{2dr}) into Equations (\ref{lm1}) and (\ref{lm2}), together with Equations (\ref{mBH1}) and (\ref{mBH2}), gives the emission luminosities due to magnetic dipole radiation,
\begin{eqnarray}
L_{\rm MDR,1} & = & \frac{3\times7^2\times2^9}{5^4\kappa^4}\left(\frac{R_{\rm S,\bullet}}{r}\right)^{12}\nonumber\\
& & \times \left(\frac{R_{\rm S,*}}{r}\right)^4\left(\frac{R_{\rm *}}{r}\right)^6a_\bullet^4cR_{\rm S,\bullet}^2B_{\rm s,*}^2\label{lmdr1},
\end{eqnarray}
and
\begin{eqnarray}
L_{\rm MDR,2} & = & \frac{13^2\times2^4}{75\kappa^5}\left(\frac{M_*}{M}\right)^4\left(\frac{R_{\rm S,\bullet}}{r}\right)^{11}\nonumber\\
& & \times \left(\frac{R_{\rm S,*}}{r}\right)^4\left(\frac{R_{\rm *}}{r}\right)^6a_\bullet^2cR_{\rm S,\bullet}^2B_{\rm s,*}^2\label{lmdr2}.
\end{eqnarray}

\subsection{Electric Dipole Radiation}
Recently \cite{Deng18} and \cite{Zhang19} found that electric dipole radiation is more significant than magnetic dipole radiation for an initially and constantly charged BH--BH inspiral. Owing to increasing charge and inspiral of a BH--NS binary in our model, the luminosity due to electric dipole radiation is obtained by
\begin{eqnarray}
L_{\rm EDR} & = & \frac{2}{3c^3}\left|\frac{d^2(Q_{\rm W}{\bf r}_{\rm \bullet})}{dt^2}\right|^2=\frac{2(Q_{\rm W}r_{\rm \bullet})^2}{3c^3}\Lambda\label{le},
\end{eqnarray}
where
\begin{eqnarray}
\Lambda=\frac{16\Omega^2\dot{r}^2}{r^2}+\left(\frac{6\dot{r}^2}{r^2}-\frac{2\ddot{r}}{r}-\Omega^2\right)^2\label{lambda}.
\end{eqnarray}
Since the second term in Equation (\ref{lambda}) gives the ratio
\begin{eqnarray}
\frac{6\dot{r}^2-2r\ddot{r}}{r^2\Omega^2} & = & \frac{3\times2^9}{25\kappa}\left(\frac{R_{\rm S,\bullet}}{r}\right)^3\left(\frac{R_{\rm S,*}}{r}\right)^2\label{ratio2},
\end{eqnarray}
we see that this ratio is close to unity for $r\sim R_{\rm S,\bullet}$, $M_{\rm *}\simeq1.4M_\odot$, and $M_{\rm \bullet}\sim 10M_\odot$. This implies $\Lambda\sim16\Omega^2\dot{r}^2/r^2$, and thus from Equations (\ref{lmdr1}) and (\ref{le}), we get
\begin{eqnarray}
\frac{L_{\rm EDR}}{L_{\rm MDR,1}}\simeq\frac{5^2\kappa^3}{3^2\times7^2\times2}\frac{1}{a_\bullet^2}\left(\frac{r}{R_{\rm S,\bullet}}\right)^7\label{ratio3}.
\end{eqnarray}
The coefficient in Equation (\ref{ratio3}) is $0.028\kappa^3$, so that the ratio, $L_{\rm EDR}/L_{\rm MDR,1}$, is much less than unity for $a_\bullet\gtrsim0.5$ and $r\sim R_{\rm S,\bullet}$. Contrary to \cite{Deng18} and \cite{Zhang19} for a constantly charged BH--BH inspiral, therefore, we find that the electric dipole radiation luminosity is negligible as compared with the magnetic dipole radiation for an increasingly charged BH--NS inspiral.

\subsection{Magnetic Reconnection}
As shown by \cite{Wald74}, if the BH spin aligns with the NS's magnetic field (${\bf B_*}$) at the position of the BH (in which case the BH spin is antiparallel to the NS's magnetic axis), then positive charges released from infinity will be accreted onto the BH along the pole and negative charges will be repelled. In this case, rotation of the charged BH along its spin axis will lead to a magnetic dipole field, whose direction at the pole is also antiparallel to that of the NS. Such a ``BH pulsar'' -- NS binary \citep{Levin18} is similar to Case 1 (an ``antiparallel'' NS--NS binary) of Figure 1 in \cite{Wang18}. An important result of this structure is that the directions of magnetic field lines in between the BH and NS are opposite to each other and a magnetic reconnection region would occur. The magnetic reconnection luminosity is estimated by \citep{Wang18}
\begin{eqnarray}
L_{\rm REC}\simeq \frac{[B_{\rm *}(r_i)]^2}{8\pi}\frac{V}{P}\label{lrec1},
\end{eqnarray}
where $r_i=r/(1+\epsilon^{1/3})$ is the magnetic reconnection distance to the NS, $\epsilon=|{\bf m}_{\rm \bullet,1}|/m_{\rm *}$ is the ratio of the magnetic dipole moments of the BH and NS (where $m_{\rm *}=B_{\rm s,*}R_{\rm *}^3$ is the NS's magnetic dipole moment), and $V\simeq (2\pi r_i)(|\dot{r}_i|P)h$ is the volume of the magnetic reconnection region with $h\simeq 0.77r_i$ being the reconnection height. From Equation (\ref{mBH1}), we find $\epsilon=6.5\times 10^{-3}a_\bullet^2(M_{\rm \bullet}/10M_\odot)^3(r/10^7{\rm cm})^{-3}\ll 1$ and $r_i\simeq r$. Thus,
we obtain the magnetic reconnection luminosity
\begin{eqnarray}
L_{\rm REC}\simeq \frac{0.31}{\kappa}\left(\frac{R_{\rm S,*}}{r}\right)\left(\frac{R_{\rm *}}{r}\right)^6cR_{\rm S,\bullet}^2B_{\rm s,*}^2\label{lrec2}.
\end{eqnarray}
It should be noted that because the magnetic dipole moment of the BH is much less than that of the NS, the luminosity $L_{\rm REC}$ is independent on the BH's spin.

The case discussed above is that the BH spin aligns with the NS's ${\bf B_*}$. The other case is that the BH spin antialigns with the NS's ${\bf B_*}$, in which case the BH spin is parallel to the NS's magnetic axis. Thus, the BH will accumulate more and more negative charges as the binary inspirals. A resultant magnetic dipole field of the BH in between the two compact objects is still antiparallel to that of the NS. In this case, a similar magnetic reconnection event would also occur at the inspiral stage.

\section{Order-of-Magnitude Estimates}\label{est}
To obtain order-of-magnitude estimates of EM emission, we here consider a BH--NS binary example, in which $M_{\rm \bullet}=10M_\odot$ and $M_{\rm *}=1.4M_\odot$. From Equations (\ref{lmdr1}), (\ref{lmdr2}), (\ref{ratio3}), and (\ref{lrec2}), we get the peak EM emission luminosities due to magnetic dipole radiation, electric dipole radiation, and magnetic reconnection at $r=r_{\rm min}$,
\begin{eqnarray}
L_{\rm MDR,1}^{\rm peak}\simeq 2.8\times 10^{43}a_\bullet^4m_{\rm *,30}^2\left(\frac{30\,{\rm km}}{r_{\rm min}}\right)^{22}\,{\rm erg}\,{\rm s}^{-1}\label{lmdr1p},
\end{eqnarray}
\begin{eqnarray}
L_{\rm MDR,2}^{\rm peak}\simeq 2.1\times 10^{39}a_\bullet^2m_{\rm *,30}^2\left(\frac{30\,{\rm km}}{r_{\rm min}}\right)^{21}\,{\rm erg}\,{\rm s}^{-1}\label{lmdr2p},
\end{eqnarray}
\begin{eqnarray}
L_{\rm EDR}^{\rm peak}\simeq 5.3\times 10^{41}a_\bullet^2m_{\rm *,30}^2\left(\frac{30\,{\rm km}}{r_{\rm min}}\right)^{15}\,{\rm erg}\,{\rm s}^{-1}\label{ledrp},
\end{eqnarray}
and
\begin{eqnarray}
L_{\rm REC}^{\rm peak}\simeq 1.8\times 10^{43}m_{\rm *,30}^2\left(\frac{30\,{\rm km}}{r_{\rm min}}\right)^7\,{\rm erg}\,{\rm s}^{-1}\label{lrecp},
\end{eqnarray}
where $m_{\rm *,30}=m_{\rm *}/10^{30}\,{\rm G}\,{\rm cm}^3$. Therefore, we conclude that magnetic dipole radiation due to the increasingly charged BH's spin and magnetic reconnection in between the BH and the NS could be most significant at the final inspiral stage of a BH--NS binary. Of course, all of the luminosities are highly dependent on $r_{\rm min}$ with different indices. They change in order as $r_{\rm min}$ increases.

We next calculate the total EM emission energies due to these energy dissipation mechanisms through $E^{\rm tot}\simeq\int_\infty^{r_{\rm min}}(L/\dot{r})dr$ and Equations (\ref{1dr}) and (\ref{lmdr1p})-(\ref{lrecp}) as follows:
\begin{eqnarray}
E_{\rm MDR,1}^{\rm tot}\simeq 6.2\times 10^{38}a_\bullet^4m_{\rm *,30}^2\left(\frac{30\,{\rm km}}{r_{\rm min}}\right)^{18}\,{\rm erg}\label{Emdr1},
\end{eqnarray}
\begin{eqnarray}
E_{\rm MDR,2}^{\rm tot}\simeq 4.9\times 10^{34}a_\bullet^2m_{\rm *,30}^2\left(\frac{30\,{\rm km}}{r_{\rm min}}\right)^{17}\,{\rm erg}\label{Emdr2},
\end{eqnarray}
\begin{eqnarray}
E_{\rm EDR}^{\rm tot}\simeq 1.9\times 10^{37}a_\bullet^2m_{\rm *,30}^2\left(\frac{30\,{\rm km}}{r_{\rm min}}\right)^{11}\,{\rm erg}\label{Eedr},
\end{eqnarray}
and
\begin{eqnarray}
E_{\rm REC}^{\rm tot}\simeq 2.4\times 10^{39}m_{\rm *,30}^2\left(\frac{30\,{\rm km}}{r_{\rm min}}\right)^3\,{\rm erg}\label{Emdr2}.
\end{eqnarray}
From these estimates, we still find that magnetic dipole radiation due to the BH's spin and magnetic reconnection could be most significant at the final inspiral stage.

In fact, the estimated energies will initially propagate in the form of a Poynting flux-dominated wind. Similar to a pulsar and a gamma-ray burst, such a wind would possibly undergo an internal gradual magnetic dissipation process \citep{Ben17,Xiao17,Xiao19,Xiao18}, leading to high-energy emission (e.g., X-ray). Detailed studies show that the conversion efficiency of the Poynting flux luminosity to X-ray luminosity is $\eta_x\sim 10^{-3}-10^{-2}$ \citep{Xiao17,Xiao19,Xiao18}. A higher value of $\eta_x$ was suggested by taking a more efficient internal energy dissipation process into account \citep{Zhang13}. According to Equation (\ref{lmdr1p}), therefore, a wind with an X-ray luminosity $L_x=\eta_xL_w\gtrsim 10^{42}\,{\rm erg}\,{\rm s}^{-1}$ at the farthest distance (i.e., detection horizon in the X-ray band), $D_L\simeq 0.1(L_w/10^{42}\,{\rm erg}\,{\rm s}^{-1})^{1/2}(F_\gamma/10^{-12}\,{\rm erg}\,{\rm cm}^{-2}\,{\rm s}^{-1})^{-1/2}\,$Gpc, would be detected by an X-ray satellite with a sensitivity of order $F_\gamma\sim 10^{-12}\,{\rm erg}\,{\rm cm}^{-2}\,{\rm s}^{-1}$.

The BH--NS merger rate density ${\cal R}_{\rm BHNS}$ remains unknown but it could be in the range of the BH--BH merger rate density ${\cal R}_{\rm BBH}$ to the NS--NS merger rate density ${\cal R}_{\rm BNS}$. The observations by aLIGO/Virgo show ${\cal R}_{\rm BBH}=53.2^{+58.5}_{-28.8}\,{\rm Gpc}^{-3}\,{\rm yr}^{-1}$ \citep{Abbott18c} and ${\cal R}_{\rm BNS}=1540^{+3200}_{-1220}\,{\rm Gpc}^{-3}\,{\rm yr}^{-1}$ \citep{Abbott17a}. Thus, the BH--NS merger rate detected by an X-ray satellite with a sensitivity of $F_\gamma$ is estimated to be in the range of $(4\pi/3)D_L^3{\cal R}_{\rm BHNS}\simeq 0.22^{+0.25}_{-0.12}(D_L/0.1\,{\rm Gpc})^3\,{\rm yr}^{-1}$ to $6.50^{+13.0}_{-5.11}(D_L/0.1\,{\rm Gpc})^3\,{\rm yr}^{-1}$. This rate is encouraging.


\section{Conclusions}\label{con}
In this paper, we have explored EM emission from the inspiral of a binary composed of a BH and an NS. At present this is one of the most interesting topics because of possible multimessenger signals from such a merger event. However, current observations of BH--BH mergers and galactic X-ray binaries indicate that the detected BHs are usually more massive than $\sim7M_\odot$. If a BH--NS binary has such a massive BH, then during its inspiral the NS whole will inevitably plunge into the BH. As a result, no EM emission is expected. Here we have assumed that the BH is rapidly rotating (i.e., $a_\bullet\gtrsim0.5$) and the NS is strongly magnetized (i.e., $B_{\rm s,*}\gtrsim10^{12}\,$G). In this case, the BH, which is immersed in the NS's magnetic field, will accumulate more and more charges during the inspiral based on the charging scenario of \cite{Wald74}. We discussed three energy dissipation mechanisms to emit EM signals: magnetic dipole radiation, electric dipole radiation, and magnetic reconnection. We found that magnetic dipole radiation due to the spin of the increasingly charged BH and magnetic reconnection in between the BH and the NS could be most significant at the final inspiral stage of the binary. Therefore, these mechanisms would give rise to a detectable EM emission signal (e.g., a short-duration X-ray transient).

\acknowledgments
The author would like to thank Xiang-Yu Wang, Xue-Feng Wu, Yun-Wei Yu, and Bing Zhang for their helpful comments and suggestions. This work was supported by the National Key Research and Development Program of China (grant No. 2017YFA0402600) and the National Natural Science Foundation of China (grant Nos. 11573014 and 11833003).

\end{document}